\documentstyle[preprint,aps]{revtex}
\begin{document}
\draft
\title{Comparison of Chiral Perturbation Theory and QCD Sum Rule
\\ Results for Pseudoscalar Isoscalar-Isovector Mixing \\
\begin{flushright}
ADP-95-2/T169 \\
hep-ph/9504252
\end{flushright}
}
\author{Kim Maltman\cite{byline}}
\address{Department of Mathematics and Statistics, York University, 4700 Keele
St., \\ North York, Ontario, Canada M3J 1P3}
\date{\today}
\maketitle
\begin{abstract}
The forms of the neutral, non-strange pseudoscalar propagator matrix and mixed
axial current correlator, $\langle 0\vert T(A_\mu^3 A_\nu^8)\vert 0\rangle$,
are discussed at next-to-leading (one-loop) order in chiral perturbation
theory, and the results compared to those obtained using QCD sum rules.
This comparison provides a check of the truncations employed in the sum
rule treatment of the current correlator.  Values for the slope of the
correlator with $q^2$ in the two approaches are found to differ by more
than an order of magnitude and the source of this discrepancy is shown
to be the incorrect chiral behavior of the sum rule result.
\end{abstract}
\pacs{12.39.Fe, 24.85.+p, 21.30.+y, 21.45.+v}

In the traditional treatment of charge symmetry breaking (CSB) in the
meson-exchange framework, contributions to CSB observables arising from
isoscalar-isovector meson mixing are obtained under the assumption that
the strength of meson mixing is $q^2$-independent.  A number of recent
papers\onlinecite{ref1,ref2,ref3,ref4,ref5,ref6,ref7,ref8,ref8new,ref8prime} ,
however,
have demonstrated that this assumption is
suspect.  Although the quantity in question (the off-diagonal element
of the matrix propagator) is the element of an off-shell Green function
and, as such, not in general invariant under allowable field redefinitions
(in the sense of Haag's theorem), nonetheless, the existence of
$q^2$-dependence for any particular choice of interpolating field
throws the standard treatment of CSB into question.  Of
special interest, among the papers mentioned above, is the treatment
of $\rho -\omega$ and $\pi -\eta$ mixing using QCD
sum rules \onlinecite{ref7,ref8}
since, without having made any apparent explicit choice
for the meson interpolating fields, the authors claim to
extract the
leading $q^2$-dependence of the off-diagonal propagator element.  If
true, this result would be extremely interesting, suggesting
that at least the leading $q^2$-dependence, was field reparametrization
independent, and hence to be incorporated in all treatments of CSB.

In this paper we critically investigate this claim in the pseudoscalar
sector (cf. Ref.~\onlinecite{ref8}~) and come to two main
conclusions.  The first is that it is not possible to extract
the $q^2$-dependence of the off-diagonal propagator matrix element
from that of the off-diagonal element of the axial vector current
correlator matrix.  That this is true in general is a consequence
of the well-known behavior of quantum field theories under allowed field
redefinitions \onlinecite{refhaag} :  S-matrix elements are
unchanged, but off-shell Green functions are not.  The off-diagonal
element of the propagator is not a physical observable; the
axial vector current correlator is. They cannot, therefore,
in general, be related.
The second, and more useful, conclusion is that chiral perturbation
theory (ChPT) provides formal constraints useful for investigating
the reliability of the trunctations employed in the sum rule treatment
of the axial vector current correlator.  The reason this is true
is that ChPT, being constructed solely via symmetry arguments,
provides the most general possible representation of the physics
of the pseudoscalar Goldstone bosons realizing the exact
symmetries of QCD and breaking the approximate chiral symmetries in
exactly the way they are broken in QCD.  The off-diagonal
element of the axial current correlator, therefore, has a low-energy
expansion in terms of current quark masses, momenta and the low-energy
constants of the effective chiral Lagrangian,  the terms of
which can be calculated in a reliable
and systematic manner.  The only approximations enter when one
truncates this expression to a given order in the chiral expansion.
However, even if the expansion converges slowly for a given
observable (typically, expansion to one-loop order for
$SU(3)_L\times SU(3)_R$ is sufficient, but a small number of
observables are known for which this is not the case -- for a general
discussion see the recent review by Ecker \onlinecite{refecker}~)
the formal dependence on the current quark masses and momenta obtained
to a given order is a rigorous consequence of QCD.  For the case at hand,
namely the off-diagonal element of the axial current correlator,
we will see below that, while the leading chiral behavior of the
$q^2$-independent part is correctly reproduced by the truncations
employed in Ref.\onlinecite{ref8} , the leading chiral
behavior of the $q^2$-dependent piece is not.  The sum
rule result for the slope with respect to $q^2$ of this correlator,
which differs numerically from the one-loop ChPT expression by
more than an order of magnitude, thus cannot be correct.

The remainder of the paper is organized as follows.  We first
revisit the attempt to relate the off-diagonal
propagator and axial current correlator matrix elements, fixing
our notation in the process.  A general expression for the
meson pole contributions to the axial current correlator is then
given in terms of the meson decay constants and two
isospin breaking parameters describing the couplings of the axial
currents $A^3_\mu$ and $A^8_\mu$ to the physical $\eta$ and
$\pi^0$, respectively.  The calculation of these
parameters is then reviewed, the development providing, in
addition, an explicit realization of the source of error in the
treatment of the relation of the propagator and correlator matrix
elements in Ref. \onlinecite{ref8} .  Finally, we compare the
one-loop ChPT result for the off-diagonal element of the axial
correlator with that obtained  via the sum rule analysis.
Using the known dependences of the physical meson masses on the
current quark masses we show that the truncations employed in the
sum rule treatment of Ref. \onlinecite{ref8} remove the leading
chiral behavior of the slope with respect to $q^2$,
and hence are unsuitable for use in treating this
feature of the correlator.

We begin with some notation.
Let $\pi$, $\eta$
represent the physical mixed-isospin $\pi^o$ and $\eta$ fields,
and $\pi_3$, $\pi_8$ the pure $I=1,\ 0$ flavor octet neutral fields.
Then one has, in general, two mixing angles, $\theta_\pi$ and $\theta_\eta$,
such that, to ${\cal O} (\theta_{\pi ,\eta})$,
\begin{eqnarray}
&&\pi =\pi_3 + \theta_\pi \pi_8\qquad \pi_3 = \pi -\theta_\pi \eta
\nonumber \\
&&\eta = -\theta_\eta \pi_3 +\pi_8\qquad \pi_8 = \theta_\eta\pi +\eta\ .
\label{one}
\end{eqnarray}
(There is, in general, $q^2$-dependent mixing so that
$\theta_\pi\not= \theta_\eta$.)  To ${\cal O}(\theta_{\pi ,\eta})$, i.e.,
to first order in isospin breaking, one defines the isospin-breaking
parameter, $\theta (q^2)$, by
\begin{equation}
\Pi_{38}(q^2)= i\int d^4x \ e^{iq.x}\langle 0\vert T(\pi_3(x)\pi_8(0))
\vert 0\rangle
\equiv {\theta (q^2)\over (q^2-m_\pi^2)(q^2-m_\eta^2)}\label{two}
\end{equation}
where, from Eqns.~(1),
$\theta (q^2) = q^2(\theta_\eta -\theta_\pi ) + (m_\eta^2\theta_\pi
-m_\pi^2\theta_\eta )\ $.
The
axial current correlator, $\Pi^{38}_{\mu\nu}$, is similarly defined via
\begin{equation}
\Pi_{\mu\nu}^{38}= i\int d^4x\ e^{iq.x}
\langle 0\vert T(A^3_\mu (x)A^8_\nu (0))\vert 0\rangle
\equiv \Pi^{38}_1(q^2)q_\mu q_\nu +\Pi^{38}_2(q^2)g_{\mu\nu} \label{three}
\end{equation}
where
$A^3_\mu$, $A^8_\nu$ are the $3,8$ members of the axial current
octet $A^a_\mu =\bar q\gamma_\mu\gamma_5{\lambda^a\over 2}q$.
In Ref. \onlinecite{ref8} the authors evaluate this correlator using
QCD sum rules, and attempt to determine $\theta (q^2)$
by considering the pseudoscalar pole contributions to the
form factor $\Pi^{38}_1$.  Invoking the PCAC relations
$\langle 0\vert A^3_\mu\vert
\pi_3 (q)\rangle = if_\pi q_\mu$ and
$\langle 0\vert A^8_\mu\vert
\pi_8 (q)\rangle = if_\eta q_\mu$ and, evaluating $\Pi_1^{38}$
in the pole approximation, they write
\begin{equation}
\Pi_1^{38}= f_\pi f_\eta\left[ {\theta_\eta\over
q^2-m_\eta^2} - {\theta_\pi\over q^2-m_\pi^2}\right]
={f_\pi f_\eta\theta (q^2)\over (q^2-m_\pi^2)(q^2-m_\eta^2)}\ .\label{four}
\end{equation}
which would allow the extraction of $\theta (q^2)$ if
$\Pi^{38}_1$ were known (e.~g. from the
sum rules treatment).
The expression (4), however, is
incorrect. The $\pi$ pole term, corresponding to the second term in
Eqn.\ (4), arises from taking
\begin{eqnarray}
\langle 0\vert A^3_\mu \vert \pi (q)\rangle &&\equiv\langle 0\vert A^3_\mu\vert
(\pi_3+\theta_\pi\pi_8)(q)\rangle =if_\pi q_\mu\nonumber \\
\langle 0\vert A^8_\mu \vert \pi (q)\rangle &&\equiv\langle 0\vert A^8_\mu\vert
(\pi_3+\theta_\pi\pi_8)(q)\rangle =if_\eta\theta_\pi q_\mu\label{five}
\end{eqnarray}
and the first term, corresponding to the $\eta$ pole, from
\begin{eqnarray}
\langle 0\vert A^8_\mu \vert \eta (q)\rangle &&\equiv
\langle 0\vert A^8_\mu\vert
(\pi_8-\theta_\eta\pi_3)(q)\rangle =if_\eta q_\mu\nonumber \\
\langle 0\vert A^3_\mu \vert \eta (q)\rangle &&\equiv
\langle 0\vert A^3_\mu\vert
(\pi_8-\theta_\eta\pi_3)(q)\rangle =-if_\pi\theta_\eta q_\mu\ .\label{six}
\end{eqnarray}
The second relations in Eqns.\ (5), (6)
are, however, false.
Although $\langle 0\vert A^3_\mu\vert\pi_8\rangle =$
$\langle 0\vert A^8_\mu\vert \pi_3\rangle = 0$ to leading order in
the chiral expansion, beyond leading order, both matrix elements
are non-zero.  This is inevitable at some order, given that
${\cal L}_{QCD}$ contains a $\Delta I=1$ piece.  The result that the
mixed-isospin current matrix element vanishes at leading order is a
consequence of chiral symmetry, and does not persist
beyond this order.  Note, moreover, that, at leading order, $\pi_3-\pi_8$
mixing is $q^2$-independent.  Thus, at this order,
$\theta_\pi =\theta_\eta$ and the $q^2$-dependence of $\theta (q^2)$
vanishes.  To determine the $q^2$-dependence, one must
go beyond leading order, but, beyond leading order, the second of
relations (5), (6) are not valid.

The corrected version of the pole approximation
to $\Pi^{38}_1$ follows from the relations
\begin{eqnarray}
&&\langle 0\vert A^3_\mu\vert\pi\rangle = if_\pi q_\mu\qquad
\langle 0\vert A^8_\mu\vert\pi\rangle = if_\pi\epsilon_1 q_\mu\nonumber \\
&&\langle 0\vert A^8_\mu\vert\eta\rangle = if_\eta q_\mu\qquad
\langle 0\vert A^3_\mu\vert\eta\rangle = -if_\eta\epsilon_2 q_\mu\ .
\label{seven}
\end{eqnarray}
The isospin-breaking parameters $\epsilon_{1,2}$ have been computed to
next-to-leading order in the chiral expansion by Gasser and Leutwyler
\onlinecite{ref9} (the calculation is reviewed below).  In Eqn.~(7),
$f_\pi$ is the physical $\pi^o$ decay constant (which differs from that
for the charged pions only at
${\cal O}\left( (m_d-m_u)^2\right)$\onlinecite{ref9} ,
and $f_\eta$ the physical $\eta$ decay constant.  ChPT implies
$f_\eta /f_\pi\simeq 1.3$\onlinecite{ref9}~.  To this (next-to-leading)
order, $\epsilon_{1,2}$, $f_\pi$ and $f_\eta$ are all independent
of $q^2$.  From Eqn.\ (7) one obtains, in the pole approximation,
\begin{equation}
\Pi_1^{38}=\Biggl[ {\epsilon_2 f_\eta^2\over q^2
-m_\eta^2} - {\epsilon_1 f_\pi^2\over q^2-m_\pi^2}\Biggr]
= \Biggl[ {q^2(\epsilon_2f_\eta^2-\epsilon_1f_\pi^2)
+(m_\eta^2\epsilon_1f_\pi^2-m_\pi^2\epsilon_2f_\eta^2)\over
(q^2-m_\eta^2)(q^2-m_\pi^2)}\Biggr] \ . \label{eight}
\end{equation}
At leading order $\epsilon_1$, $\epsilon_2$, $\theta_\pi$ and
$\theta_\eta$ are all equal to $\theta_0$, the leading-order
$\pi_3 -\pi_8$ mixing angle arising from the leading order
off-diagonal term in the meson mass-squared matrix,
\begin{equation}
\theta_0 = {\sqrt{3} (m_d-m_u)\over 4 (m_s -\hat m)}\label{nine}
\end{equation}
where ${\hat m} =(m_u+m_d)/2$.  Also at leading order $f_\pi =f_\eta =F$
so that, at this order, the numerator in Eqn.\ (8) reduces to
$F^2 \theta_0$, which is identical to the leading order
expression for $f_\pi f_\eta \theta (q^2)$.  The expression (4) is
thus correct at leading order in the chiral expansion.
Beyond leading order, however, $\epsilon_2f_\eta^2\not= f_\pi f_\eta
\theta_\eta$ and
$\epsilon_1f_\pi^2\not= f_\pi f_\eta \theta_\pi$ (see below), and knowing the
numerator in Eqn.\ (8) no longer allows one to obtain that of Eqn.\ (3).
In particular, the crucial $q^2$-dependence of $\theta (q^2)$,
which arises only at next-to-leading order and beyond, is completely
inaccessible from the $q^2$-dependence of the numerator of (8).

We now provide explicit expressions for all quantities
appearing in the above discussion, at
one-loop order in the chiral expansion,
obtained using the
low-energy effective chiral Lagrangian, ${\cal L}_{eff}$,
of Ref.\ \onlinecite{ref9}~.
All notation is as defined there.
We also keep terms
only to first order in isospin
breaking, i.e., drop terms of ${\cal O}\left( (m_d-m_u)^2\right)$.

Let us first discuss the meson propagator matrix, for the standard
choice of meson fields \onlinecite{ref9}~.
$\pi^{(0)}_3$ , $\pi^{(0)}_8$
are the bare $\pi_3$ , $\pi_8$ fields.
The bare propagator matrix, to one-loop order, is of the form
\begin{equation}
\Delta (q) =
\left(
\begin{array}{cc}
Z_3^{-1}(q^2-m_3^2)&\ell_{38} \\
\ell_{38}&Z_8^{-1}(q^2-m_8^2)
\end{array}\right)^{-1}\label{ten}
\end{equation}
where $Z_3$, $Z_8$ and $m_3^2$, $m_8^2$ are the
$\pi^{(0)}_3$ , $\pi^{(0)}_8$
one-loop wavefunction renormalization constants and squared-masses
in the isospin symmetry limit.  Expressions for
$m_3^2$ and $m_8^2$ in terms of the parameters of ${\cal L}_{eff}$
are given in Ref.\ \onlinecite{ref9} ,
and those for $Z_3$, $Z_8$ and $\ell_{38}$
are easily computable from ${\cal L}_{eff}$.  In what follows we
write $Z_3^{-1}=1+z_3$, $Z_8^{-1}=1+z_8$ and
$\ell_{38}=\ell_{(0)}+\ell^{(1)0}+\ell^{(1)1}q^2$.
$\ell_{(0)}$, $\ell^{(1)0}$ and $\ell^{(1)1}$ are
$q^2$-independent and ${\cal O}(m_d-m_u)$.
$\ell_{(0)}$ is of leading chiral order and
$z_3$, $z_8$, $\ell^{(1)0}$, $\ell^{(1)1}$ next-to-leading order.
$m_3^2$ and $m_8^2$ contain both leading and
next-to-leading pieces.  The renormalized $\pi_3$, $\pi_8$
fields, $\pi_t^r$ and $\pi_e^r$, are obtained, as usual, by
\begin{eqnarray}
&&\pi_t^r=(Z^{-1/2})_{33}\pi^{(0)}_3 +(Z^{-1/2})_{38}\pi^{(0)}_8 \nonumber \\
&&\pi_e^r=(Z^{-1/2})_{83}\pi^{(0)}_3 +(Z^{-1/2})_{88}\pi^{(0)}_8 \label{eleven}
\end{eqnarray}
where $Z$ is the wavefunction renormalization matrix.  From Eqn.\ (10)
one has
\begin{equation}
Z^{-1/2}=\left(
\begin{array}{cc}
1+{1\over 2}z_3&{1\over 2}\ell^{(1)1} \\
{1\over 2}\ell^{(1)1}&1+{1\over 2}z_8\end{array}\right)\label{twelve}
\end{equation}
and from this one obtains the renormalized propagator matrix,
to ${\cal O}(m_d-m_u)$,
\begin{equation}
\Delta^{(r)}(q)=\left(
\begin{array}{cc}
q^2-m_3^2&-m_{te}^2 \\
-m_{te}^2&q^2-m_8^2\end{array}\right)^{-1}\label{thirteen}
\end{equation}
where $m_{te}^2=-[\ell_{(0)}+\ell^{(1)0}+{1\over 2}\ell^{(1)1}(m_3^2+m_8^2)
-{1\over 2}\ell_{(0)}(z_3+z_8)]$.
Using the explicit expressions for $z_3$, $z_8$, $\ell_{(0)}$,
$\ell^{(1)0}$ and $\ell^{(1)1}$, one may verify that $m_{te}^2$
is now finite, and diagonalize the mass-squared matrix to find
the physical, renormalized $\pi^0$, $\eta$ fields.  Writing
\begin{eqnarray}
&&\pi^0=\pi_t^r +\theta_r\pi_e^r \nonumber \\
&&\eta =-\theta_r\pi_t^r+\pi_e^r\label{fourteen}
\end{eqnarray}
with $\theta_r=-m_{te}^2/(m_8^2-m_3^2)$, one finds, explicitly,
\begin{eqnarray}
\theta_r=&&\theta_0\Biggl[ 1-3\mu_\pi +2\mu_K +\mu_\eta
+\left({3m_8^2+m_3^2\over 64\pi^2F^2}\right)
\left( 1+{m_\pi^2\over {\bar m}_K^2-m_\pi^2}
\log (m_\pi^2/{\bar m}_K^2)\right) \nonumber \\
&&\mbox{}-{32B_0(m_s-\hat m)\over F^2}(3L_7^r+L_8^r)\Biggr]\equiv
\theta_0(1+\delta\theta_r)\label{fifteen}
\end{eqnarray}
where $\mu_P=\left[ m_P^2\log (m_P^2/\mu^2)\right] / 32\pi^2F^2$ with $\mu$
the (dimensional regularization) renormalization scale,  $L_k^r$ are
the renormalized low-energy constants as defined in Ref.\ \onlinecite{ref9} ,
${\bar m}_K^2$ is the average $K$ mass-squared and $F$, $B_0$ are
the two constants appearing in the lowest order part of the chiral
Lagrangian ($F=f_\pi$ in leading order).  The expression (14)
is not yet of use to us since the $\pi_t^r$ and $\pi_e^r$ fields
have mixed isospin.  We may re-write (14) as
\begin{eqnarray}
&&\pi^0=(1+{1\over 2}z_3)(\pi^{(0)}_3+{\hat\theta}_1\pi^{(0)}_8)\nonumber \\
&&\eta =(1+{1\over 2}z_8)(-{\hat\theta}_2\pi^{(0)}_3+\pi^{(0)}_8)
\label{sixteen}
\end{eqnarray}
where
\begin{eqnarray}
{\hat\theta}_1=&&\theta_0\Biggl[ 1-{7\over 3}\mu_\pi +{4\over 3}\mu_K +\mu_\eta
+\left( {3m_\eta^2+m_\pi^2\over 64\pi^2F^2}\right)
\left( 1+{m_\pi^2\over {\bar m}_K^2-m_\pi^2}
\log (m_\pi^2/{\bar m}_K^2)\right) \nonumber \\
&&\mbox{}+\left({m_\eta^2-m_\pi^2\over 64\pi^2F^2}\right)
\left( 1+\log ({\bar m}_K^2/\mu^2)\right)
-{32({\bar m}_K^2-m_\pi^2)\over F^2}(3L_7^r+L_8^r)\Biggr]\equiv
\theta_0(1+\delta{\hat\theta}_1)\nonumber \\
{\hat\theta}_2=&&\theta_0\Biggl[ 1-{11\over 3}\mu_\pi +
{8\over 3}\mu_K +\mu_\eta
+\left({3m_\eta^2+m_\pi^2\over 64\pi^2F^2}\right)
\left( 1+{m_\pi^2\over {\bar m}_K^2-m_\pi^2}
\log (m_\pi^2/{\bar m}_K^2)\right)\nonumber \\
&&\mbox{}-\left({m_\eta^2-m_\pi^2\over 64\pi^2F^2}\right)
\left( 1+\log ({\bar m}_K^2/\mu^2)\right)
-{32({\bar m}_K^2-m_\pi^2)\over F^2}(3L_7^r+L_8^r)\Biggr]\equiv
\theta_0(1+\delta{\hat\theta}_2)\ .\label{seventeen}
\end{eqnarray}
The $\delta{\hat\theta}_{1,2}$ are next-to-leading order quantities.
If $\ell_{38}$ had had no $q^2$-dependence, we would have
had ${\hat\theta}_1={\hat\theta}_2$ at this point.  To compare
with Eqn.\ (1) we recast the above results in terms of the
isospin-pure renormalized fields, $\pi^{(0)r}_3$ and $\pi^{(0)r}_8$,
i.e., the renormalized fields in the isospin symmetry limit,
$\pi^{(0)r}_3\equiv (1+{1\over 2}z_3)\pi^{(0)}_3$ and
$\pi^{(0)r}_8\equiv (1+{1\over 2}z_8)\pi^{(0)}_8$, obtaining
\begin{eqnarray}
&&\pi^0=\pi^{(0)r}_3+\theta_0(1+\delta{\hat\theta}_1+{1\over 2}z_3
-{1\over 2}z_8)\pi^{(0)r}_8\nonumber \\
&&\eta =-\theta_0(1+\delta{\hat\theta}_2+{1\over 2}z_8 -{1\over 2}z_3)
\pi^{(0)r}_3+\pi^{(0)r}_8\ .\label{eighteen}
\end{eqnarray}
Thus, formally, we find $\theta_\pi =\theta_0(1+\delta{\hat\theta}_1+
{1\over 2}z_3-{1\over 2}z_8)$,
$\theta_\eta =\theta_0(1+\delta{\hat\theta}_2+{1\over 2}z_8 -{1\over 2}z_3)$.

Let us turn to the axial current correlator, $\Pi_{\mu\nu}^{38}$.
Although the low-energy representation of the product $A_\mu^3A_\nu^8$
contains contact terms, these do not affect $\Pi^{38}_1$ to
next-to-leading order, so we may restrict our
attention to the product of representations of the individual
currents.  These are
easily read off from ${\cal L}_{eff}$ of
Ref.\ \onlinecite{ref9} , with the result
\begin{eqnarray}
A_\mu^3 =&& -F\partial_\mu\pi^{(0)}_3 - {16B_0\over F}(m_s+2{\hat m})\, L_4
\partial_\mu\pi^{(0)}_3 -{16B_0\over F}{\hat m}\, L_5\partial_\mu
\pi^{(0)}_3 \nonumber \\
&&\mbox{}+{1\over 3F}(4\pi^+\pi^- + K^+K^- +K^0{\bar K}^0)
\partial_\mu\pi^{(0)}_3\nonumber \\
&&+{8B_0(m_d-m_u)\over \sqrt{3} F}L_5\partial_\mu\pi^{(0)}_8
+ {1\over \sqrt{3} F}(K^+K^- - K^0{\bar K}^0)\partial_\mu\pi^{(0)}_8
+\cdots \nonumber \\
A_\mu^8 =&& -F\partial_\mu\pi^{(0)}_8 - {16B_0\over F}(m_s+2{\hat m})\, L_4
\partial_\mu\pi^{(0)}_8 -{16B_0\over 3F}(2m_s+{\hat m})\, L_5\partial_\mu
\pi^{(0)}_8 \nonumber \\
&&\mbox{}+{1\over F}(K^+K^- +K^0{\bar K}^0)
\partial_\mu\pi^{(0)}_8\nonumber \\
&&+{8B_0(m_d-m_u)\over \sqrt{3} F}L_5\partial_\mu\pi^{(0)}_3
+ {1\over \sqrt{3} F}(K^+K^- - K^0{\bar K}^0)\partial_\mu\pi^{(0)}_3
+\cdots \label{nineteen}
\end{eqnarray}
where the $+\cdots$ indicates terms higher order in the fields, and other
terms which do not contribute to $\Pi_1^{38}$ to one-loop order.
As claimed earlier, one sees, beyond the leading order terms
$-F\partial_\mu\pi^{(0)}_3$ in $A_\mu^3$ and $-F\partial_\mu\pi^{(0)}_8$
in $A_\mu^8$, terms which will couple $A_\mu^3$ to $\pi_8$ and
$A_\mu^8$ to $\pi_3$.  From the expressions (19)
for $A_\mu^{3,8}$ and the inverted form of the relations
(16), i.e.,
\begin{eqnarray}
&&\pi^{(0)}_3=(1-{1\over 2}z_3)\pi^0-\theta_0(1+\delta{\hat\theta}_1
-{1\over 2}z_8)\eta\nonumber \\
&&\pi^{(0)}_8=(1-{1\over 2}z_8)\eta +\theta_0(1+\delta{\hat\theta}_2
-{1\over 2}z_3)\pi^0\label{twenty}
\end{eqnarray}
(valid to ${\cal O}(m_d-m_u)$ ,
and to next-to-leading order in the chiral expansion)
one may easily verify the results of Gasser and Leutwyler for
$f_\pi$, $f_\eta$, $\epsilon_1$ and $\epsilon_2$.
The following equivalent forms display explicitly the problems with
the second of Eqns.~(5), (6):
\begin{eqnarray}
&&\langle 0\vert A_\mu^3\vert\eta (q)\rangle
\equiv -i\, f_\eta\epsilon_2 q_\mu = -i\, f_\pi q_\mu\left(
\theta_\eta +c_{12}+{8B_0(m_d-m_u)\over \sqrt{3} F^2}L_5
+{({\hat L}_{K^+}-{\hat L}_{K^0})\over \sqrt{3}\, F^2}
\right)\label{twentyone}\\
&&\langle 0\vert A_\mu^8\vert\pi^0(q)\rangle
\equiv i\, f_\pi\epsilon_1 q_\mu = i\, f_\eta q_\mu\left(
\theta_\pi -c_{12}-{8B_0(m_d-m_u)\over \sqrt{3} F^2}L_5
-{({\hat L}_{K^+}-{\hat L}_{K^0})\over \sqrt{3}\, F^2}\right)\label{twentytwo}
\end{eqnarray}
where ${\hat L}_P$ is the tadpole loop integral
${\hat L}_P=\mu^{4-d}\int {d^dk\over (2\pi )^d}{i\over (k^2-m_P^2)}$
and $c_{12}=\theta_0 (\delta{\hat\theta}_1-\delta{\hat\theta}_2+z_3-z_8)$.
The first terms on the right hand sides of Eqns.~(21), (22) are the only ones
retained in the treatment of Ref.\ \onlinecite{ref8}~.  The remaining
terms are present due to the direct couplings of
$\pi^{(0)r}_8$ (the leading piece of $\eta$) to $A_\mu^3$,
at ${\cal O}(m_d-m_u)$, and of $\pi^{(0)r}_3$ (the leading piece of
$\pi^0$) to $A_\mu^8$, at ${\cal O}(m_d-m_u)$.
As already expected on general principles, the second of
Eqns.\ (5), (6) are seen explicitly to be incorrect at
next-to-leading order as, in consequence, is Eqn.\ (4), which is based
on them:  the isospin-breaking parameter, $\theta (q^2)$, of
the pseudoscalar propagator matrix cannot be extracted directly from
the mixed axial current correlator.

It is important to stress that this conclusion is inescapable, independent
of whether or not the chiral expansions for $\epsilon_{1,2}$ are
well-converged at one-loop order.  The $q^2$-dependences of
$\Pi^{38}_1$ and $\theta (q^2)$ arise only at one-loop, and not
at leading, order in the chiral expansion.  Eqns.~(21), (22) demonstrate
that terms of the same order as those kept must be dropped for
the $q^2$-dependent piece of the numerator of Eqn. (8) to reduce to
$f_\pi f_\eta \theta (q^2)$.  Since such a procedure would be
inconsistent (moreover the additional terms are not small, numerically),
we confirm that the general argument, based on the behavior of
quantum field theories under field redefinitions, is not accidentally
evaded in the pseudoscalar sector.

Let us now turn to the more interesting point of using the chiral
expansion as a constraint on the truncations employed in the
sum rule treatment of the axial current correlator.  For this we
will give the explicit expressions for $\Pi^{38}_{1,2}$ to one-loop
in ChPT and compare the formal quark mass dependences of these
expressions with those of the results of the sum rule analysis.  The
latter may be obtained from the results of Ref. \onlinecite{ref8}
(written in terms of the physical $\pi^0$, $\eta$ masses) by using the
known leading- plus next-to-leading expansions for $m_{\pi ,\eta}$
in terms of the quark masses \onlinecite{ref9} .  It is important
to remember that the axial current correlator is a physical
object.  As such, its dependence on $q^2$ is {\bf not} to be identified with
the $q^2$-dependence of the angle which diagonalizes the inverse
propagator matrix (as obtained in Refs.~\onlinecite{ref2,ref6,ref7}~)~.
That angle, like $\theta (q^2)$ (though distinct from it), is
dependent on the choice of meson fields.

The low-energy representation
of the product of axial currents, $A_\mu^3A_\nu^8$,
consists of two parts, (1) the product of the low-energy representations
of the individual currents, as given in Eqns.~(19), and (2) a set of
contact terms associated with those terms in ${\cal L}_{eff}$ quadratic
in the external axial fields.  To next-to-leading order, explicit
calculation shows that the product of the representations of the
individual currents contributes only to $\Pi_1^{38}$, and the
contact terms only to $\Pi_2^{38}$.  For $\Pi_2^{38}$ one finds
\begin{equation}
\Pi_2^{38}={B_0(m_d-m_u)\over \sqrt{3}}\Biggl[ {3\over 32\pi^2}
\left(\log (\bar m_K^2/\mu^2)+1\right) -8L_5^r\Biggr]\ .\label{neweqn}
\end{equation}
Although not discussed in detail in Ref.\ \onlinecite{ref8} ,
the pole approximation
employed there leads to a $\Pi_2^{38}$
(as defined here) which is proportional to $q^2$ and hence vanishes
as $q^2\rightarrow 0$.  It follows, from Eqn.\ (23), that
$\Pi_2^{38}$ cannot be properly modelled using only pseudovector
meson pole terms, as in Ref.\ \onlinecite{ref8}~.  Turning to the remaining
form factor, $\Pi_1^{38}$, we note that, to this order in the
chiral expansion, the pole approximation is exact.  The correct one-loop
expression for $\Pi_1^{38}$ is, therefore, that already written
down in Eqn.~(8), with $f_\pi$, $f_\eta$, $\epsilon_1$ and $\epsilon_2$
the one-loop expressions of
Ref.\ \onlinecite{ref9}~.
In obtaining numerical results below
we have rescaled the values of $\epsilon_{1,2}$ quoted in
Ref.\ \onlinecite{ref9} by
a factor of $1.22$, in order to account for the larger-than-expected
violations of Dashen's theorem\onlinecite{ref10,ref11,ref12} .  We thus have
as input $\epsilon_1=1.67\times 10^{-2}$, $\epsilon_2=1.35\times 10^{-2}$,
$f_\pi =93$ MeV, and $f_\eta /f_\pi =1.3$.  Then, setting
$g_\eta =f_\eta^2\epsilon_2$ and $g_\pi =f_\pi^2\epsilon_1$ and
rewriting Eqn.\ (8) in the form
\begin{equation}
\Pi_1^{38}=\left( {g_\eta\over q^2-m_\eta^2}
-{g_\pi\over q^2-m_\pi^2}\right)
\ =\left( {q^2(g_\eta -g_\pi )+(g_\pi m_\eta^2 -g_\eta m_\pi^2)
\over (q^2 -m_\eta^2)(q^2-m_\pi^2)}\right)\label{twentythree}
\end{equation}
to facilitate comparison with Ref.\ \onlinecite{ref8} , we find
$g_\eta = 197$ MeV$^2$,
$g_\pi = 144$ MeV$^2$,
to be compared with the values $g_\eta =143$ MeV$^2$, $g_\pi =139$ MeV$^2$
extracted via the sum rule analysis \onlinecite{ref8}~.
(These values have been rescaled from those quoted in Ref. \onlinecite{ref8}
 by the same factor of $1.22$ discussed above.)
To leading order in the Borel mass, $M_B$ of
Ref.\ \onlinecite{ref8} ,
\begin{equation}
g_\pi =g_\eta ={m_u\langle 0\vert {\bar u}u\vert 0\rangle
-m_d\langle 0\vert {\bar d}d\vert 0\rangle \over \sqrt{3} (m_\eta^2-m_\pi^2)}
\label{twentyfive}
\end{equation}
which, since $\langle 0\vert {\bar u}u\vert 0\rangle
=\langle 0\vert {\bar d}d\vert 0\rangle =-F^2B_0$ to leading order in
the chiral expansion, implies (again to leading order)
$g_\pi =g_\eta =\theta_0\, F^2\ $,
the leading order chiral result.  The procedure of truncating
the operator product expansion (OPE) for $A_\mu^3A_\nu^8$ at
operators of dimension 6 and dropping terms of ${\cal O}(m_q^2)$
employed in Ref.\ \onlinecite{ref8} thus correctly reproduces the leading
order chiral constraints.  The interesting $q^2$-dependence of $\Pi_1^{38}$,
however, enters first only at next-to-leading order in the chiral
expansion and here, comparing the results of the two approaches for
the coefficient, $g_\eta -g_\pi$, of $q^2$ in the numerator of
the last expression for $\Pi^{38}_1$ in
Eqn.\ (24), ChPT to one loop predicts $g_\eta -g_\pi =53$ MeV$^2$,
whereas the results of Ref.\ \onlinecite{ref8} , rescaled for the violations
of Dashen's theorem, produce $g_\eta -g_\pi =3.9$ MeV$^2$, more
than an order of magnitude smaller.  If we consider the leading chiral
behavior implicit in the sum rule result we can easily discover the source of
the discrepancy.  Using
the expressions from Ref.\ \onlinecite{ref8} to evaluate
$g_\eta -g_\pi$, one obtains, to ${\cal O}(m_d-m_u)$,
\begin{equation}
g_\eta -g_\pi ={(m_d-m_u)\over \sqrt{3}}\langle 0\vert {\bar q}q\vert
0\rangle \left( {(m_\eta^2+m_\pi^2)\over 2M_B^4} + {\cal O}(1/M_B^6)\right)
\label{twentyseven}
\end{equation}
where $\langle 0\vert {\bar q}q\vert
0\rangle =\langle 0\vert {\bar u}u\vert
0\rangle\simeq\langle 0\vert {\bar d}d\vert
0\rangle$, $M_B$ is the Borel mass and the higher order terms
are also higher order in the meson masses (and hence higher order in
the chiral expansion).
Using the known chiral expansions of the meson masses, one obtains,
for the leading term in the chiral expansion of $g_\eta -g_\pi$,
\begin{equation}
g_\eta -g_\pi =\theta_0 F^2\left( {8\over 9}{B_0^2(m_s-{\hat m})
(m_s+2{\hat m})\over M_B^4} +\cdots \right)\label{twentyeight}
\end{equation}
where $+\cdots$ represents terms higher order in the chiral expansion.
In contrast, making the same expansion for $g_\eta =f_\eta^2
\epsilon_2$ and $g_\pi =f_\pi^2\epsilon_1$ using the relevant expressions
from
Ref.\ \onlinecite{ref9} , one finds
\begin{equation}
g_\eta -g_\pi =\theta_0F^2\left( {(m_\pi^2-{\bar m}_K^2)\over 8\pi^2F^2}
\log ({\bar m}_K^2/\mu^2) -{B_0(m_s-{\hat m})\over 8\pi^2F^2}
+{32B_0(m_s-{\hat m})\over 3F^2}L_5^r +\cdots \right)\ .\label{twentynine}
\end{equation}
The result of Eqn.\ (27) begins at ${\cal O}(m_q^2)$, that of Eqn.\ (28) at
${\cal O}(m_q)$ (plus the ubiquitous chiral logs):
the truncation scheme of Ref.\ \onlinecite{ref8} has thus removed the
leading contribution to the slope of $\Pi_1^{38}$, which
accounts for the smallness of the numerical result.  It would be
of interest to determine which aspect of the truncation
scheme is the source of the problem.
The situation here appears analogous to
that encountered in the sum rules analysis
of the nucleon mass, where the usual approximate treatment
of the continuum contributions to the phenomenological side of the
sum rule leads to violations of the known chiral behavior of $M_N$ with
${\hat m}$, and only a more careful analysis restores the correct
next-to-leading order behavior\onlinecite{ref13,ref14} .

In summary, we have demonstrated (1) that the mixing parameter, $\theta (q^2)$,
of the pseudoscalar meson propagator matrix cannot be extracted, except
at leading order in the chiral expansion (where it is independent of $q^2$),
from a treatment of the mixed axial correlator $\Pi_{\mu\nu}^{38}$, and
(2) that the QCD sum rules treatment of $\Pi_{\mu\nu}^{38}$, in the
approximation of dropping terms of ${\cal O}(m_q^2)$ and truncating the
OPE to operators of dimension 6 or less, fails completely at next-to-leading
order.  Since the pole approximation for
$\Pi_1^{38}$ breaks down
beyond next-to-leading order, the sum rule method appears unlikely
to provide any information about the $q^2$-dependence of
$\Pi_{\mu\nu}^{38}$ beyond that already
known, at next-to-leading order, from the chiral expansion, unless the
multi-meson continuum is taken into account in the phenomenological
side of the sum rule analysis.  One could,
however, profitably use the results of ChPT to provide useful
constraints on the application of the sum rules
method to few-body systems.

\acknowledgements

The hospitality of the Department of Physics and Mathematical Physics of
the University of Adelaide and the continuing financial support of the
Natural Sciences and Research Engineering Council of Canada are gratefully
acknowledged.


\begin{references}
\bibitem[*]{byline}Current address: Department of Physics and Mathematical
Physics, University of Adelaide, Adelaide, South Australia 5005, Australia
\bibitem{ref1}T.~Goldman, J.~A.~Henderson and A.~W.~Thomas,
Few-Body Systems {\bf 12} (1992) 123
\bibitem{ref2}K.~Maltman, Phys.\ Lett.\ {\bf B313} (1993) 203
\bibitem{ref3}G.~Krein, A.~W.~Thomas, and A.~G.~Williams, Phys.\ Lett.\
{\bf B317} (1993) 293
\bibitem{ref4}J.~Piekarewicz and A.~G.~Williams, Phys.\ Rev.\ {\bf C47}
(1993) R2461
\bibitem{ref5}J.~Piekarewicz, Phys.\ Rev.\ {\bf C48} (1993) 1555
\bibitem{ref6}K.~Maltman and T.~Goldman, Nucl.\ Phys.\ {\bf A572} (1994) 682
\bibitem{ref7}T.~Hatsuda, E.~M.~Henley, T.~Meissner and G.~Krein,
Phys.\ Rev.\ {\bf C49} (1994) 452
\bibitem{ref8}C.~-T.~Chan, E.~M.~Henley and T.~Meissner, ``$\pi -\eta$
Mixing from QCD Sum Rules'', Univ. Washington preprint, 1994
\bibitem{ref8new}H.~B.~O'Connell, B.~C.~Pearce, A.~W.~Thomas and
A.~G.~Williams, Phys.\ Lett.\ {\bf B336} (1994) 1
\bibitem{ref8prime}K.~L.~Mitchell, P.~C.~Tandy, C.~D.~Roberts and
R.~T.~Cahill, Phys.\ Lett.\ {\bf B335} (1994) 282
\bibitem{refhaag}R.~Haag, Phys. Rev. {\bf 112} (1958) 669
\bibitem{refecker}G.~Ecker, ``Chiral Perturbation Theory'', Universitat
Wien preprint UWThPh-1994-49, to appear in Progress in Particle and
Nuclear Physics {\bf 35}
\bibitem{ref9}J.~Gasser and H.~Leutwyler, Nucl.\ Phys.\ {\bf B250} (1985) 465
\bibitem{ref10}K.~Maltman and D.~Kotchan, Mod.\ Phys.\ Lett.\ {\bf A5} (1990)
2457
\bibitem{ref11}J.F.~Donoghue, B.R.~Holstein and D.~Wyler,
Phys.\ Rev.\ Lett.\ {\bf 69} (1992) 3444; Phys.\ Rev.\ {\bf D47},
(1993) 2089
\bibitem{ref12}J.~Bijnens, Phys.\ Lett.\ {\bf B306} (1993) 343
\bibitem{ref13}D.~K.~Griegel and T.~D.~Cohen, Phys.\ Lett.\ {\bf B333} (1994)
27
\bibitem{ref14}S.~H.~Lee, S.~Choe, T.~D.~Cohen and D.~K.~Griegel,
``QCD Sum Rules and Chiral Logarithms'', University
of Maryland preprint 95-067, HEP-PH-9411428, 1994
\end{references}
\end{document}